\documentclass{article}
\usepackage{graphicx} 
\usepackage{float}
\usepackage{subcaption}
\usepackage{tikz}
\usetikzlibrary{matrix, decorations.pathreplacing, positioning}
\usepackage{booktabs}
\usepackage{url}
\usepackage{multirow}

\title{Analysis of Typhoons Path Using Functional Data Analysis Methods}
\author{
Jimin Kim \thanks{Department of Statistics, Ewha Womans University, Korea. E-mail: jmpape21@gmail.com}}
\date{}

\begin{document}

\maketitle

\begin{abstract}
Accurate prediction of typhoon trajectories is essential for mitigating the impact of these extreme weather events. This study proposes a functional data analysis (FDA) framework for modeling and forecasting typhoon paths using historical trajectory data. Latitude and longitude sequences are represented as smooth functional observations, and a function-on-function regression model is employed to capture the temporal dynamics of typhoon movement. While the baseline model demonstrates reasonable performance for typical bow-shaped trajectories, it exhibits limited accuracy for non-standard paths. To address this limitation, a clustering-based extension is introduced, wherein typhoons are grouped by trajectory shape prior to regression. This two-stage approach improves predictive accuracy by enabling localized modeling adapted to structural variations in the data. The results demonstrate the practical utility of combining FDA with clustering for robust and flexible typhoon trajectory forecasting.
\end{abstract}


\section{Introduction}
Typhoons, which frequently occur in the western North Pacific and impact regions such as the Korean Peninsula, pose serious social and economic risks. Accurate trajectory prediction is therefore critical for effective disaster preparedness and risk mitigation.

Although numerical weather prediction models remain central to operational forecasting, their high computational cost and limited ability to capture the full range of trajectory variability have prompted growing interest in data-driven alternatives. Recent studies have explored machine learning and deep learning approaches for typhoon path prediction, leveraging large-scale historical data to model complex spatiotemporal patterns. \cite{giffard2020}

This study explores the use of Functional Data Analysis (FDA) as a principled framework for modeling typhoon trajectories. FDA treats each typhoon path not as a sequence of discrete points, but as a smooth function evolving over time, thereby preserving the continuous and dynamic nature of movement. By representing trajectories in this functional form, it becomes possible to apply regression models that capture temporal dependencies more effectively than conventional techniques.

A function-on-function regression model is first constructed to predict future typhoon paths using the final segments of their trajectories. Typhoons typically follow paths in which latitude increases consistently, while longitude decreases during the early westward phase and then increases as the trajectory shifts eastward. The model effectively captures this overall structure but remains limited in representing trajectories with non-monotonic longitudinal variation.

To address this limitation, a clustering-based modeling framework is introduced. Functional clustering is applied to group typhoon trajectories with similar shapes, and separate regression models are trained within each cluster. This two-stage approach allows the model to adapt to local structural differences in trajectory patterns and improves predictive performance.

The remainder of this paper is organized as follows: Section 2 provides a brief overview of Functional Data Analysis. Section 3 outlines the proposed modeling methodology, including both the baseline and clustering-based regression frameworks. Section 4 presents the experimental results, both quantitative and qualitative. Section 5 offers a discussion of the findings and concludes the study.

\section{Functional Data Analysis}

In the real world, data are observed in various forms, such as images, text, and tabular structures, each with distinct characteristics. A wide range of analytical methods have been developed to address these different types of data. Most conventional data representations are discrete, where each cell contains a single value, and are typically expressed as vectors or matrices.

For instance, consider temperature measurements collected throughout a single day. While temperature varies continuously over time, practical and technical constraints make it impossible to record values at every instant. As a result, observations are typically obtained at fixed intervals, such as every two or three hours. This illustrates a common scenario in which an inherently continuous phenomenon is represented by a finite set of discrete measurements.

Functional Data Analysis (FDA) is a statistical framework designed to address this limitation. Unlike standard approaches that treat data as discrete vectors, FDA considers each observation as a function $x_{i}(t)$ defined over a continuous domain such as time. This allows the analysis to incorporate the smooth and dynamic structure of the data.

Because it is rarely possible to collect truly functional data, FDA typically involves constructing smooth functional approximations from discrete observations. This step, referred to as functional representation, is a key component of FDA. In this approach, a smooth function is obtained by expressing each observation as a linear combination of basis functions.

Basis functions are predefined functions that span a functional space, similarly to how basis vectors span a vector space. Using this approach, each functional observation is written as follows: 

$$x_{i}(t) = \sum_{k=1}^{K}c_{ik}\phi_{k}(t)$$

where $\phi_{k}(t)$ denotes the $k-$th basis function, and $c_{ik}$ is the corresponding coefficient for the $i-$th individual. The choice of basis functions depends on the structure of the data. B-spline bases are commonly used due to their flexibility in capturing local variations, while Fourier bases are appropriate for modeling periodic patterns.

The coefficients $c_{ik}$ are estimated using the least squares method, which minimizes the difference between the observed values and the basis expansion evaluated at the same time points. The estimation is formulated as follows:

$$ \min_{c_{ik}, \cdots, c_{iK}} \sum_{j=1}^{n} \left( x_{i}(t_{j}) - \sum_{k=1}^{K}c_{ik}\phi_{k}(t_{j}) \right)^{2}$$

The resulting coefficients provide a compact representation of the functional data and serve as the foundation for the application of various FDA techniques, including functional principal component analysis (FPCA) and functional regression.

Once functional data are properly represented, a range of statistical methods can be applied within the functional framework. Among these, functional regression is one of the most widely used approaches. It extends classical regression models by allowing the predictors, the responses, or both to be functions defined over a continuous domain. 

Functional regression models are typically categorized based on the type of variables involved. The three most common formulations are as follows:

\subsection{Function-on-Scalar Regression}

The response is a function, and the predictors are scalar variables.

$$y_{i}(t) = \alpha (t) + \sum_{j=1}^{p} {z_{ij} \beta_{j}(t)} + \epsilon_{i}(t)$$

This model describes how scalar predictors influence a functional response. For example, it can be used to evaluate how variables such as gender, body weight, or treatment assignment affect a subject's temperature profile over time. The coefficient functions $\beta_{j}(t)$ capture how the effect of the predictors vary across the domain $t$.

\subsection{Scalar-on-Function Regression}

The response is a scalar variable, and the predictors are functions.

$$y_{i} = \alpha + \int {x_{i}(t) \beta(t) dt} + \epsilon_{i}$$

This model explains how a functional predictor contributes to a scalar outcome. For instance, it can be applied to model how a subject's daily blood glucose curve affects a final health score. The coefficient function $\beta(t)$ represents the time-varying contribution of the predictor to the response.

\subsection{Function-on-Function Regression}

Both the predictor and the response are functions.

$$y_{j}(s) = \alpha (s) + \int {\beta(s,t) x_{i}(t) dt} + \epsilon_{i}(s)$$

This model characterized the dynamic relationship between functional predictors and functional responses. It allows for modeling of how the value of a predictor at time $t$ affects the response at time $s$. The coefficient $\beta (s,t)$ provides a flexible representation of this interaction across domains $s$ and $t$.

Given that the objective of this study is to predict future typhoon trajectories based on past movement patterns, a function-on-function regression model is employed to model the temporal relationship between historical and future typhoon paths.

\section{Analysis Method}

\subsection{Data Description}

\begin{table}[hbt] 
    \centering
    \caption{Data Description}\vspace{0.5ex}
    {\tabcolsep=19pt
    \begin{tabular}{ll}
    \hline \hline
    Variable & Description\\ \hline
    time & Observation timestamp\\
    indicator & Typhoon indicator\\
    grade & Typhoon grade\\
    lat & Latitude (in degrees)\\
    lon & Longitude (in degrees)\\
    cenPrs & Central pressure (hPa)\\
    maxWindSpd & Maximum wind speed (knots)\\
    dirNlnhRad50, shRad30, longRad30, shtRad50 & Radii information (nautical miles)\\
    landfall & Landfall indicator\\
    \hline \hline
    \end{tabular}}
\label{tab1}\vspace{1.5ex}
\end{table}

This study employs the Best Track Data compiled by the Regional Specialized Meteorological Center (RSMC) Tokyo, covering typhoons in the western North Pacific from 1951 to 2023. The data set contains 71,088 records for 1,894 typhoons, with observations taken at 6-hour intervals. Each record includes 12 variables, such as observation time, typhoon grade, central pressure, maximum wind speed, and radii information. In this study, latitude and longitude are used to represent the trajectories.

Applying Functional Data Analysis (FDA) requires all functional observations to have the same length. However, as shown in the histogram in Figure 1, the number of recorded time points per typhoon varies widely, ranging from 2 to 111, depending on the lifespan and movement characteristics. This variability necessitates standardization of trajectory length for functional modeling.

\begin{figure}[hbt]
    \centerline{\includegraphics[width=0.8\columnwidth]{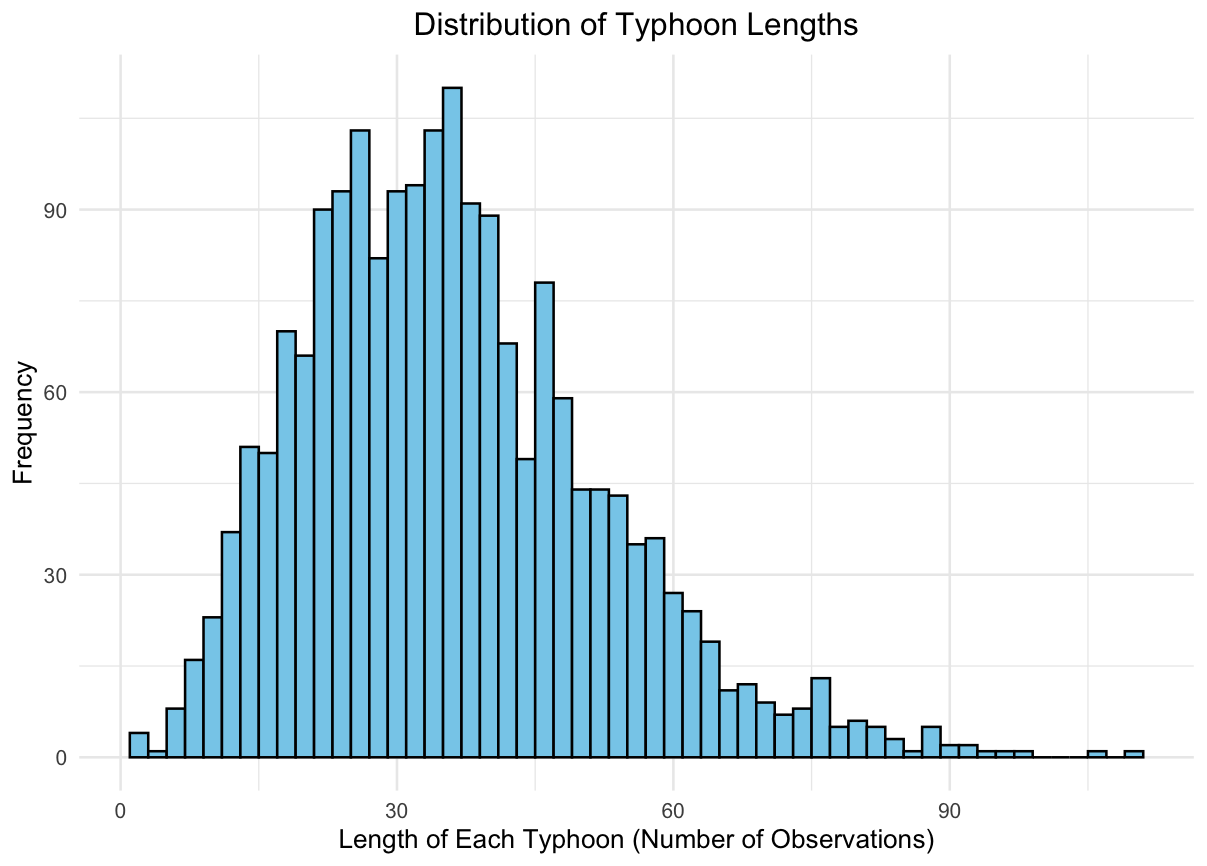}}\vspace{0.5ex}
    \caption{histogram of typhoon lifespan}
    \label{figure1}\vspace{1.5ex}
\end{figure}

To ensure consistency in the length of functional observations, only typhoons with at least 32 recorded time points (i.e., equivalent to eight days) are retained. For each selected typhoon, the last 32 observations are extracted, allowing for uniform trajectory lengths across all samples. Additional datasets with lengths of 40 and 48 are also constructed in the same manner. This preprocessing step yields consistently structured functional data, enabling direct application of FDA techniques and fair model comparisons.

\subsection{functional data analysis}

To construct a predictive model of typhoon trajectories, each path is represented as a sequence of 32 time points, corresponding to eight days of movement recorded at six-hour intervals. As illustrated in Figure 2, we divide each trajectory into two segments: the first 24 time points are designated as the predictor portion, denoted by $X$, and the remaining 8 points constitute the response portion, denoted by $Y$. This temporal division enables the modeling of how earlier movements influence the subsequent progression of the typhoon’s path.

In the resulting data matrix, each column corresponds to an individual typhoon, and each row represents a specific time point. One distinctive feature of functional data analysis , particularly when implemented in R, is that the input data must be arranged in a $p \times n$ structure where $p$ is the number of time points and $n$ is the number of functional observations. This contrasts with the conventional $n \times p$ structure used in many statistical or machine learning contexts. Accordingly, the original data matrix is transposed prior to analysis to align with this format.

\begin{figure}[htb] 
\centering
\begin{tikzpicture}
\matrix[matrix of nodes,
        nodes in empty cells,
        row sep=0.1cm,
        column sep=0.1cm,
        nodes={minimum width=1cm, minimum height=0.6cm, anchor=center},
        ampersand replacement=\&] (m) {
     \& \textcolor{gray}{obs$_1$} \& \textcolor{gray}{obs$_2$} \& $\cdots$ \& \textcolor{gray}{obs$_{885}$} \& \textcolor{gray}{obs$_{886}$} \& $\cdots$ \& \textcolor{gray}{obs$_{1107}$} \\
    \textcolor{gray}{$t_1$}   \& $x_{1,1}$ \& $x_{2,1}$ \& $\cdots$ \& $x_{885,1}$ \& $x_{886,1}$ \& $\cdots$ \& $x_{1107,1}$ \\
    \textcolor{gray}{$t_2$}   \& $x_{1,2}$ \& $x_{2,2}$ \& $\cdots$ \& $x_{885,2}$ \& $x_{886,2}$ \& $\cdots$ \& $x_{1107,2}$ \\
    \textcolor{gray}{$\vdots$} \& $\vdots$ \& $\vdots$ \& \& $\vdots$ \& $\vdots$ \& \& $\vdots$ \\
    \textcolor{gray}{$t_{24}$} \& $x_{1,24}$ \& $x_{2,24}$ \& $\cdots$ \& $x_{885,24}$ \& $x_{886,24}$ \& $\cdots$ \& $x_{1107,24}$ \\
    \textcolor{gray}{$t_{25}$} \& $x_{1,25}$ \& $x_{2,25}$ \& $\cdots$ \& $x_{885,25}$ \& $x_{886,25}$ \& $\cdots$ \& $x_{1107,25}$ \\
    \textcolor{gray}{$\vdots$} \& $\vdots$ \& $\vdots$ \& \& $\vdots$ \& $\vdots$ \& \& $\vdots$ \\
    \textcolor{gray}{$t_{32}$} \& $x_{1,32}$ \& $x_{2,32}$ \& $\cdots$ \& $x_{885,32}$ \& $x_{886,32}$ \& $\cdots$ \& $x_{1107,32}$ \\
};

\draw[decorate,decoration={brace,amplitude=6pt},thick]
  ([yshift=0pt]m-2-8.east) -- ([yshift=0pt]m-5-8.east) node[midway,xshift=1cm]{\small$\mathbf{X}$};

\draw[decorate,decoration={brace,amplitude=6pt},thick]
  ([yshift=0pt]m-6-8.east) -- ([yshift=0pt]m-8-8.east) node[midway,xshift=1cm]{\small$\mathbf{Y}$};

\draw[decorate,decoration={brace,amplitude=6pt,mirror},thick]
  ([xshift=0pt]m-8-2.south west) -- ([xshift=0pt]m-8-5.south east) node[midway,yshift=-0.5cm]{\textbf{train}};

\draw[decorate,decoration={brace,amplitude=6pt,mirror},thick]
  ([xshift=0pt]m-8-6.south west) -- ([xshift=0pt]m-8-8.south east) node[midway,yshift=-0.5cm]{\textbf{test}};

\end{tikzpicture}
\caption{Functional Data Structure of Typhoon Paths}
\end{figure}

By organizing the data in this way, each trajectory is treated as a smooth function over time, enabling the application of FDA techniques such as basis function expansion and function-on-function regression. The structure shown in Figure 2 provides a clear representation of how the predictor and response segments are defined and used in the modeling process. For model development and evaluation, the complete dataset of 1,107 typhoon trajectories is randomly partitioned into training and test sets in an 8:2 ratio.

\begin{figure}[b!]
    \centerline{\includegraphics[width=1.0\columnwidth]{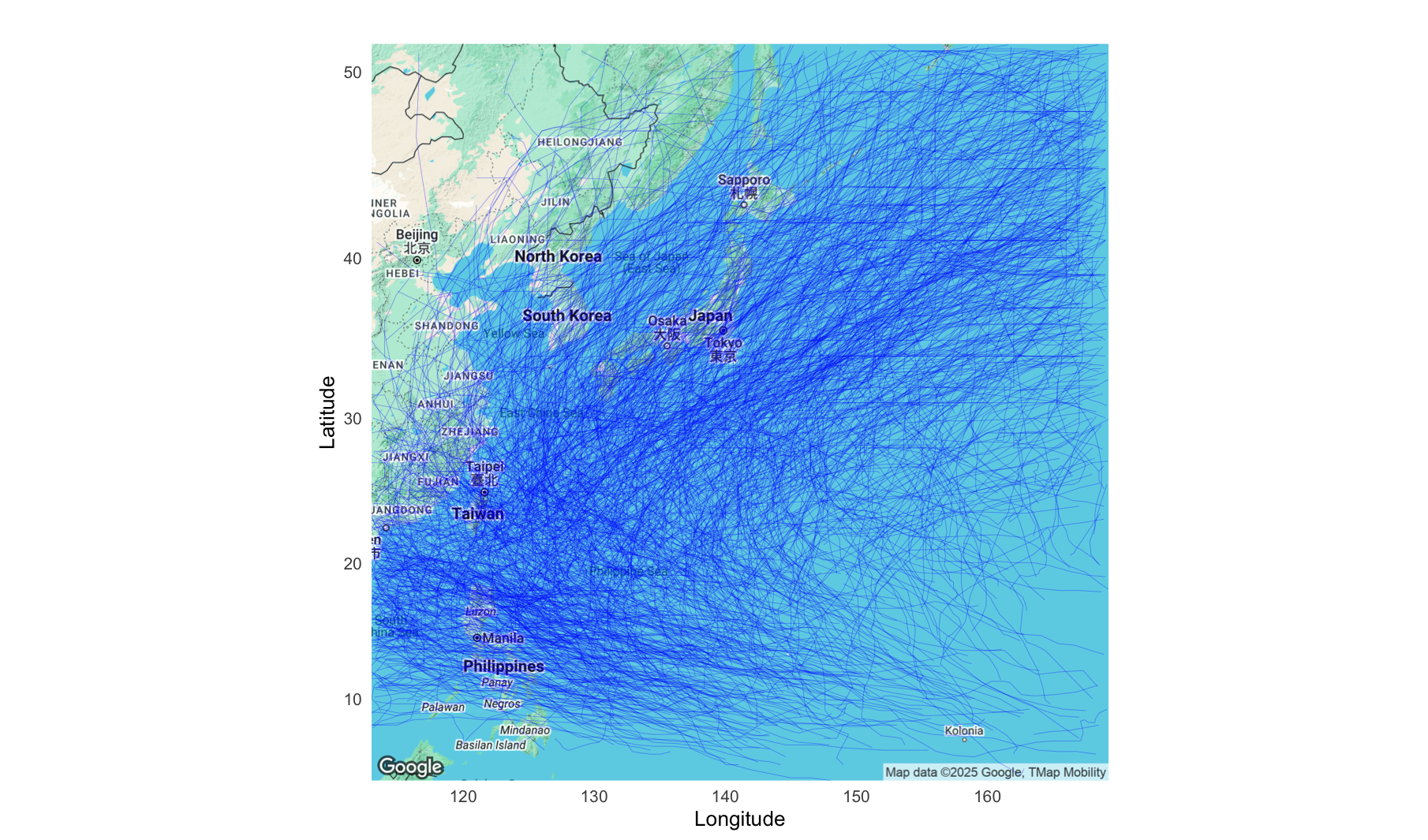}}\vspace{0.9ex}
    \caption{Typhoon Tracks}
    \label{figure3}\vspace{1.5ex}
\end{figure}

\vfill\eject
\subsection{Modeling}

The modeling approach adopted in this study begins with a functional regression framework designed to take advantage of the continuous and time-dependent nature of typhoon movement. By modeling entire trajectory curves rather than discrete points, this approach enables the capture of smooth temporal patterns in both latitude and longitude. However, while effective in identifying dominant movement trends, this method exhibits limitations when faced with the diverse and irregular paths that typhoons often take.

To address these shortcomings, we extend the baseline model by introducing an additional stage: a two-step modeling strategy that incorporates functional clustering prior to regression. By first grouping typhoons according to the shape of their trajectories, and then fitting separate regression models within each cluster, the revised framework allows the model to better reflect the heterogeneity in movement patterns. 

\subsubsection{regression without clustering}

Function-on-function regression was independently applied to latitude and longitude coordinates, using the first 24 recorded time points of each typhoon trajectory as predictors and the subsequent 8 as responses. This modeling framework leverages the continuous nature of typhoon movement and captures the underlying temporal structure of the data.

As shown in Figure 3, the majority of typhoons in the dataset follow a characteristic path: an initial northward progression, often accompanied by a slight westward drift, followed by a pronounced turn toward the northeast. This bow-shaped trajectory is especially common among typhoons originating in the western North Pacific and moving toward the East Asian region.

\begin{figure}[hbt]
    \centerline{\includegraphics[width=1.0\columnwidth]{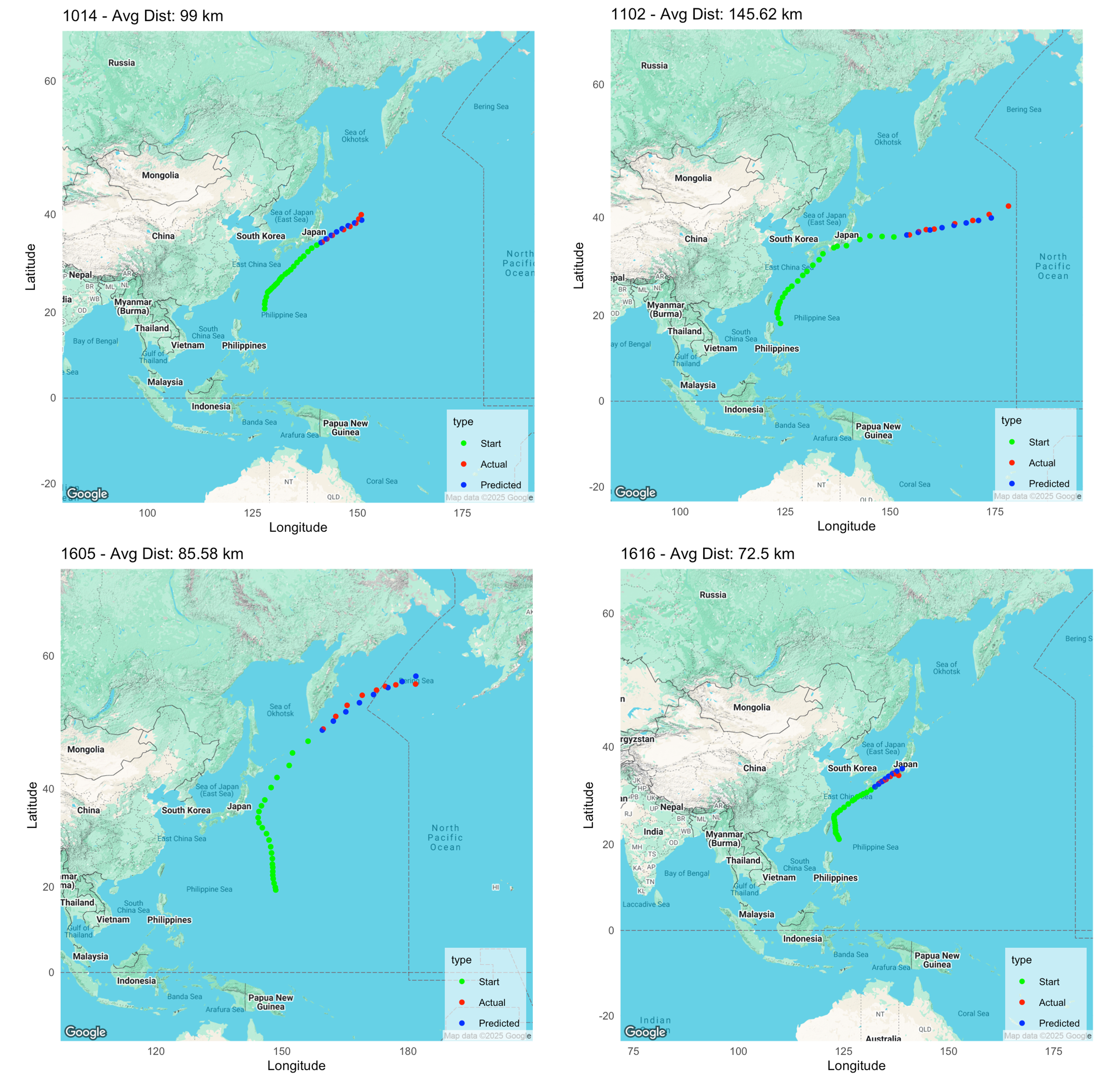}}\vspace{0.9ex}
    \caption{Good case of regression without clustering}
    \label{figure4}\vspace{1.5ex}
\end{figure}

When a given typhoon adheres to this dominant pattern, the regression model tends to perform well. Figure 4 presents an example in which the predicted path aligns closely with the observed trajectory. The four-digit number in each title indicates the year and month in which the typhoon originated. The accompanying value, labeled {\it Avg Dist}, denotes the average haversine distance between the predicted and observed points. The definition and computation of this distance metric are described in detail in Section 4.1. In this case, the model accurately captures both the directional trend and the curvature of the movement, demonstrating a strong predictive capacity under typical conditions.

However, the dominance of such standard trajectories in the training data introduces a notable modeling bias. The regression model, having captured the prevailing path pattern, tends to overgeneralize this structure, even when applied to trajectories that deviate from it. Consequently, typhoons characterized by westward movement or non-monotonic directional changes are often predicted inaccurately.

\begin{figure}[hbt]
    \centerline{\includegraphics[width=1.0\columnwidth]{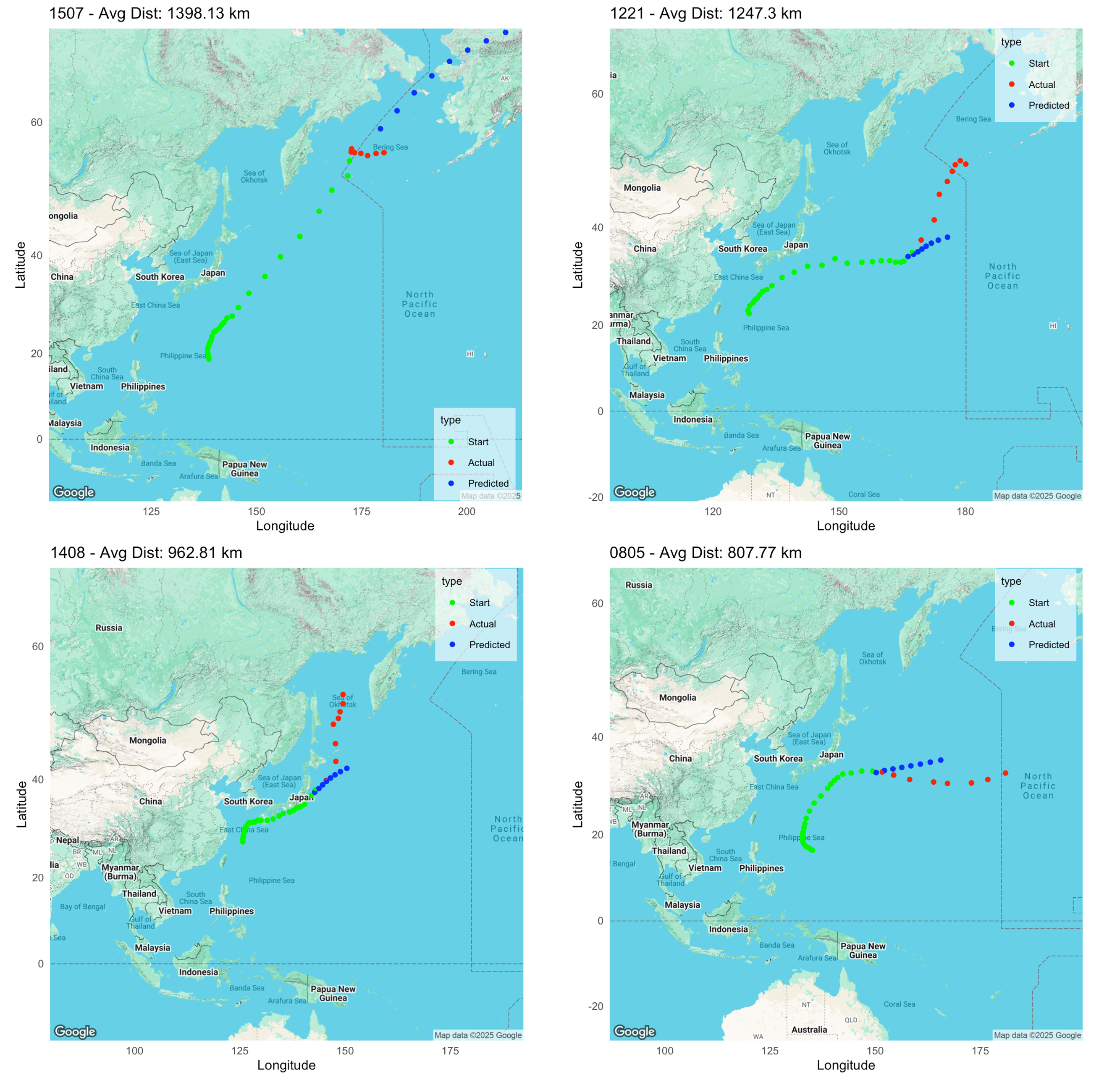}}\vspace{0.9ex}
    \caption{Bad case of regression without clustering}
    \label{figure5}\vspace{1.5ex}
\end{figure}

Figure 5 illustrates such a failure case. The predicted trajectory (in blue) diverges markedly from the observed path (in red), reflecting a tendency to default to the average northeastward trend. This example underscores the limited capacity of the model to adapt to structural variations in typhoon movement and highlights the need for a more flexible and locally adaptive modeling approach.

\subsubsection{regression with clustering}

To address the limitations of the baseline regression model discussed in Section 3.1, particularly its tendency to overfit the dominant bow-shaped trajectory pattern, a clustering-based regression framework is proposed. This method is designed to capture the structural diversity of typhoon paths by grouping similar trajectories and fitting specialized regression models within each group.

\begin{figure}[!t]
    \centerline{\includegraphics[width=1.0\columnwidth]{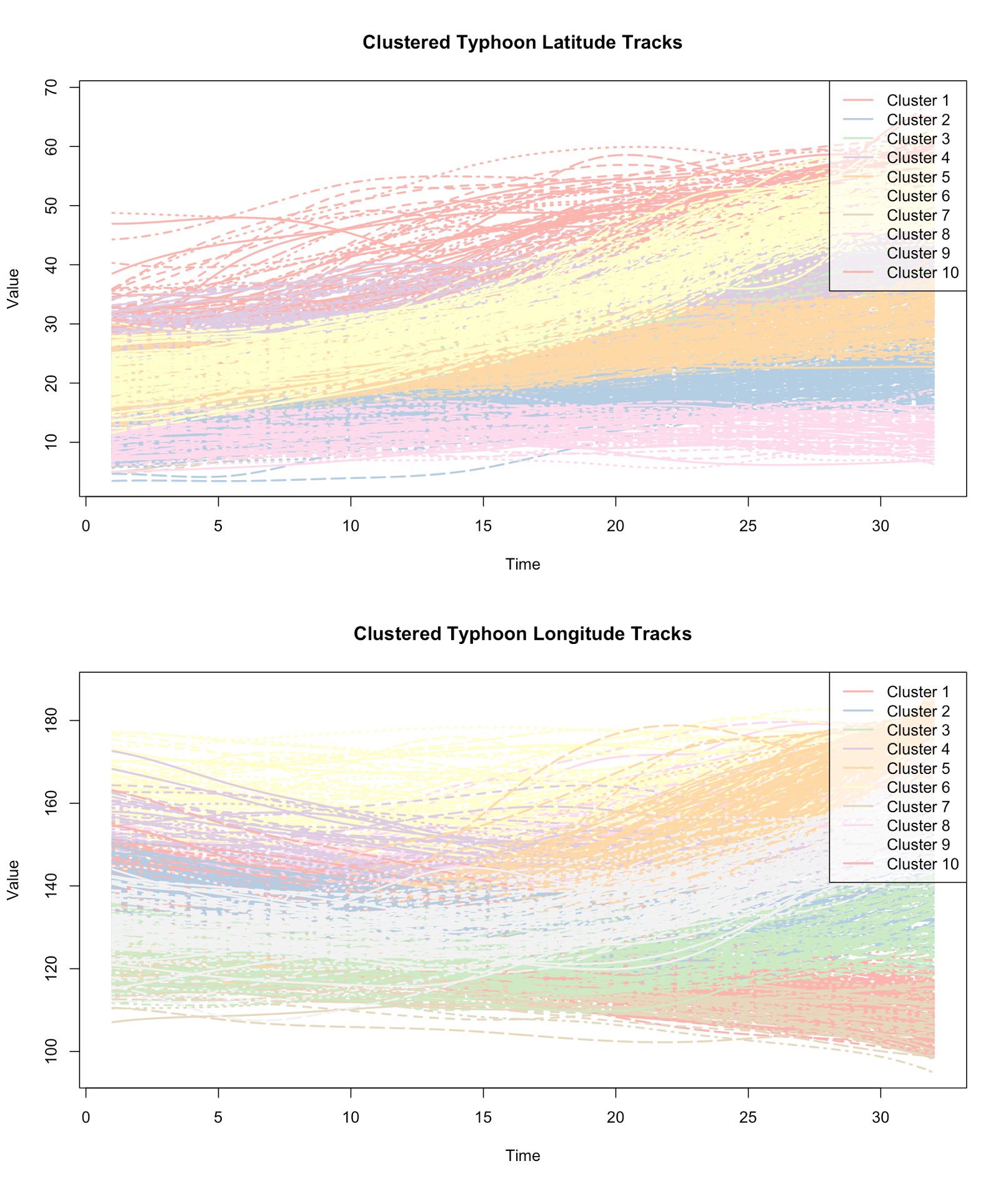}}\vspace{0.9ex}
    \caption{Results of k-means clustering}
    \label{figure6}\vspace{1.5ex}
\end{figure}

The modeling process consists of two main stages: functional clustering and cluster-specific regression. In the first stage, latitude and longitude trajectories are treated independently as functional data and clustered according to their shape. As shown in Figure 2, the $X$ segment of the training data—representing the first 24 time points—serves as the input for clustering. The k-means algorithm is employed for this purpose, and Figure 6 presents the results of applying k-means clustering with 10 clusters to each of the latitude and longitude components.

Determining the appropriate number of clusters is nontrivial: too few clusters may obscure meaningful variation, while too many may fragment the data and hinder model generalization. To address this, the number of clusters is systematically varied from 1 to 10 for both latitude and longitude, resulting in a total of 100 cluster combinations. Each combination defines a unique pairing of coordinate-based clusters.

In the second stage, a separate function-on-function regression model is trained for each of the 100 cluster pairs using the corresponding training subsets. To evaluate model performance, predictions are generated across the entire test set under each cluster configuration. The prediction error is then averaged across all test cases for each combination. The cluster pair yielding the lowest average error is selected as the optimal configuration for final model deployment.

This strategy ensures that model selection is empirically driven, rather than predetermined, and allows the regression framework to adapt to the structural characteristics of the data. By accounting for trajectory-level heterogeneity, the model offers enhanced flexibility and improved predictive accuracy.

\begin{figure}[hbt]
    \centerline{\includegraphics[width=1.0\columnwidth]{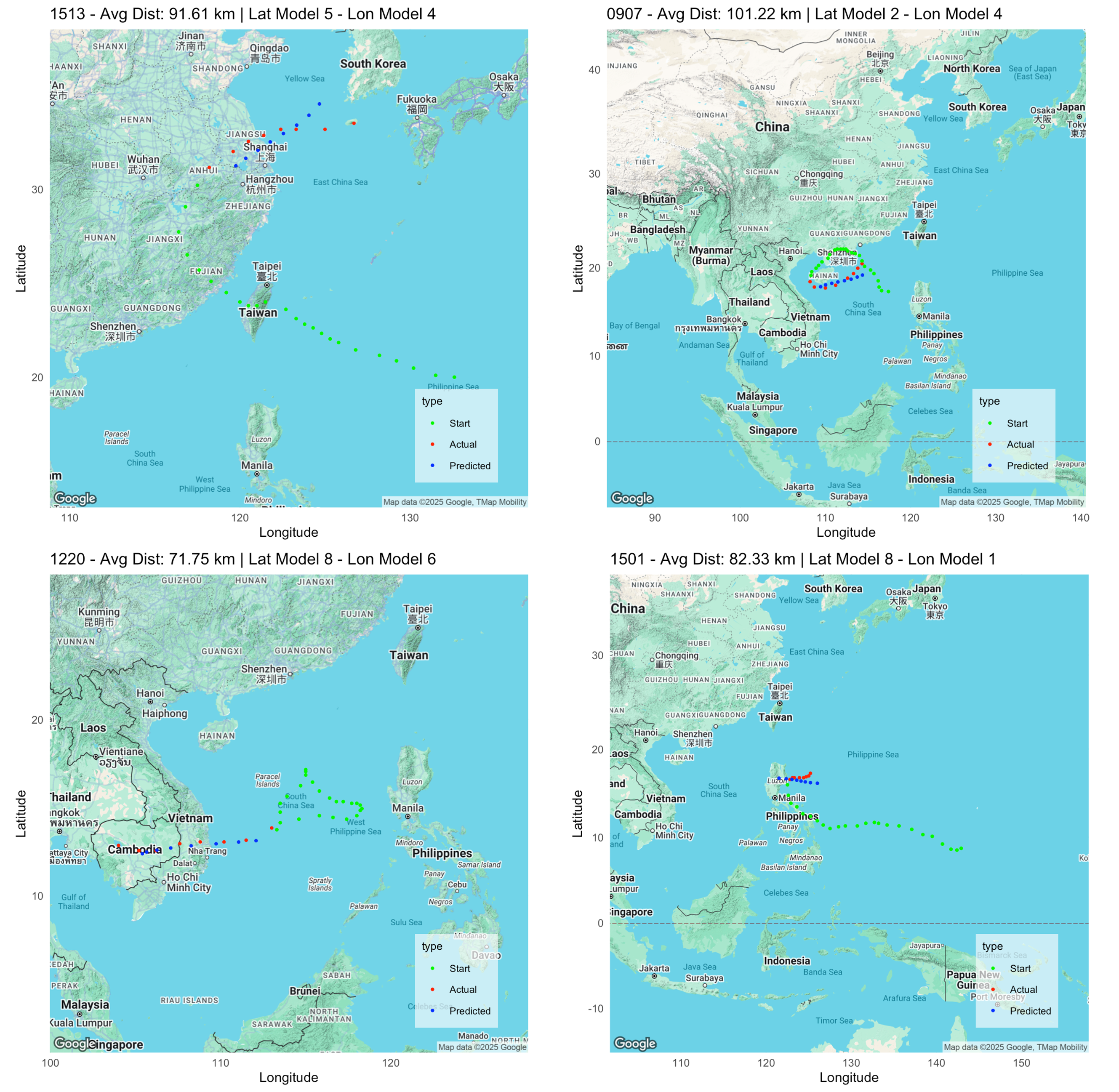}}\vspace{0.9ex}
    \caption{good case of regression with clustering}
    \label{fig}\vspace{1.5ex}
\end{figure}

Figure 7 presents a representative case where the proposed model accurately predicts a typhoon trajectory that diverges from the dominant pattern, highlighting its effectiveness in handling nonstandard movement.

\section{Results}

\subsection{Evaluation Metric}

To quantitatively assess the predictive performance of the proposed models, a distance-based error metric is employed. As there is no universally accepted standard for evaluating typhoon trajectory predictions, the error is defined as the average spatial discrepancy between the predicted and observed points in each test trajectory. Specifically, the prediction error for a given test case is calculated as the mean distance between the predicted coordinates $(\hat{lat}_{25}, \hat{lon}_{25}) \cdots (\hat{lat}_{32}, \hat{lon}_{32})$ and the corresponding ground truth $(lat_{25}, lon_{25}) \cdots (lat_{32}, lon_{32})$.

Each location is represented by a latitude–longitude pair, which must be converted into a meaningful geographic distance. To accomplish this, the haversine formula is used, which accounts for the curvature of the Earth and computes the great-circle distance between two points on a sphere based on their coordinates.

The haversine distance $d$ between two points $(lat_{1}, lon_{1})$ and $(lat_{2}, lon_{2})$ is given by the following:

$$ d = 2r \cdot \arcsin{\left( \sqrt{\sin^{2}{ \left( {\Delta \phi} \over {2} \right)} + \cos{\phi_{1}} \cos{\phi_{2}} \sin^{2}{ \left( {\Delta \lambda} \over {2} \right)}} \right)},$$

where $\phi$ and $\lambda$ denote the latitude and longitude in radius, $\Delta \phi = \phi_{2} - \phi_{1}, \Delta \lambda = \lambda_{2} - \lambda_{1}$, and $r$ is the radius of the Earth (assumed to be 6,371km).

Using this formula, the pairwise distances between the predicted and observed coordinates are computed for all eight points, and their average is reported as the final prediction error for each trajectory. This metric offers a consistent and interpretable basis for comparison across models, independent of geographic location or path shape.

\subsection{Quantitative Results}

\subsubsection{Regression without Clustering vs Regression with Clustering}

\begin{table}[hbt]
\footnotesize
\centering
\caption{Average prediction errors for the two modeling strategies}\label{tab2}\vspace{0.5ex}
{\tabcolsep=21.7pt
\begin{tabular}{lll}
\hline \hline
    \textbf{Model} & Regression without Clustering   & Regression with Clustering  \\ \hline
    \textbf{Results} & 484.2542 km & 236.977 km \\ \hline \hline
    \end{tabular}}
\captionsetup{justification=centering}
\end{table}

Table 2 presents a summary of the average prediction errors for the two modeling strategies under comparison. The reported values are averaged over 10 repeated simulations to account for variability. The baseline model, which applies a single global function-on-function regression, yields a mean haversine distance of 484.25 km. In contrast, the clustering-based model, incorporating trajectory-level grouping, reduces the error to 236.98 km.

This result demonstrates that the proposed approach reduces the average prediction error by more than 50\%, highlighting the effectiveness of localizing regression models according to path shape. The improvement is particularly meaningful given the variability of typhoon movement patterns: while the baseline model tends to regress toward a generalized northeastward path, the clustering-based model is better able to adapt to nonstandard trajectories by leveraging structurally homogeneous training subsets.

These findings clearly support the hypothesis that incorporating structural information through clustering enhances predictive accuracy. The substantial reduction in error underscores the limitations of globally trained models in capturing diverse typhoon behaviors and validates the use of cluster-specific modeling as a more flexible and robust alternative.

\subsubsection{Performance across Cluster Combinations}

To evaluate the influence of clustering granularity on model performance, latitude and longitude were independently clustered into 1 to 10 groups, producing 100 unique combinations. For each configuration, the model was trained separately, and the average prediction error across all test trajectories was computed. The results are shown in Table 3.

Prediction accuracy generally improved as the number of clusters increased, through gains diminished beyond a certain point. The lowest average error, 236.97 km, was obtained with 10 latitude clusters and 9 longitude clusters. This configuration is adopted for all subsequent analyses.

\begin{table}[htb]
\centering
\caption{Average Distance (km) by Longitude and Latitude}
\begin{tabular}{c|cccccccccc}
\hline \hline
\textbf{lon $\backslash$ lat} & k=1 & k=2 & k=3 & k=4 & k=5 & k=6 & k=7 & k=8 & k=9 & k=10 \\
\midrule
k=1 & 561.40 & 519.02 & 504.14 & 500.46 & 486.97 & 481.34 & 483.42 & 481.58 & 479.16 & 483.19 \\
k=2 & 492.77 & 447.76 & 432.53 & 427.91 & 411.45 & 406.13 & 403.19 & 403.44 & 402.53 & 406.08 \\
k=3 & 501.05 & 455.10 & 379.96 & 435.18 & 418.69 & 412.19 & 413.02 & 414.80 & 408.14 & 412.78 \\
k=4 & 444.53 & 386.30 & 378.08 & 385.36 & 354.05 & 347.26 & 346.90 & 341.47 & 335.70 & 346.77 \\
k=5 & 408.97 & 355.04 & 340.24 & 353.41 & 335.65 & 309.47 & 309.14 & 341.57 & 325.50 & 329.78 \\
k=6 & 395.94 & 342.73 & 334.68 & 352.91 & 301.22 & 294.19 & 372.29 & 337.84 & 325.55 & 306.93 \\
k=7 & 394.20 & 340.17 & 324.35 & 317.48 & 315.17 & 304.58 & 284.48 & 304.74 & 287.41 & 285.58 \\
k=8 & 378.79 & 314.70 & 308.18 & 291.61 & 271.49 & 264.44 & 291.13 & 294.09 & 273.33 & 276.21 \\
k=9 & 371.85 & 316.40 & 297.69 & 291.01 & 269.80 & 265.36 & 262.37 & 261.40 & 278.43 & 278.26 \\
k=10 & 376.95 & 305.21 & 308.94 & 299.22 & 295.85 & 289.13 & 299.61 & 280.87 & $\mathbf{236.97}$ & 277.17 \\
\hline
\bottomrule
\end{tabular}
\end{table}

To further assess the impact of input length on model performance, additional experiments were conducted using longer trajectory segments. Specifically, while the prediction $Y$ was fixed at 8 time points across all settings, the input segment $X$ was extended from 24 to 32 and 40, resulting in total sequence lengths of 32, 40, and 48, respectively. This design allowed for investigation into whether increasing the amount of past information improves the accuracy of future path prediction.

In this setup, only the clustering-based model was evaluated, and the train/test ratio was kept consistent with previous experiments. For each input length, clustering was performed on the extended $X$ segments, and prediction errors were computed across combinations of latitude and longitude clusters ranging from k=1 to k=10. 

However, as longer input sequences were required, the number of available typhoon samples decreased. For example, when the input length was set to 40, the last 40 observations of each trajectory were extracted (e.g., using the tail(40) function in R). This necessarily excluded typhoons with fewer than 40 recorded points, reducing the data size to 709 samples. Similarly, when the input length was extended to 48, only trajectories with at least 48 recorded points could be used, resulting in a further reduction of the data size to 425 samples.

\begin{table}[hbt]
\centering
\caption{Prediction errors by trajectory length and data size}
\label{tab4}
\vspace{0.5ex}
\begin{tabular}{c|ccc}
\hline \hline 
\multirow{2}{*}{\textbf{Data Size}} & \multicolumn{3}{c}{\textbf{Length}} \\
\cline{2-4}
& \textbf{32} & \textbf{40} & \textbf{48} \\
\hline 
1107 & \textbf{236.97} km (k=9, k=10) & -- & -- \\
709 & 299.81 km (k=9, k=8)  & 303.85 km (k=9, k=9) & -- \\
425 & 316.28 km (k=9, k=10) & 359.77 km (k=9, k=10) & 409.52 km (k=9, k=10) \\
\hline \hline
\end{tabular}
\end{table}

The results presented in Table~\ref{tab4} underscore a trade-off between input sequence length and data availability in the context of typhoon trajectory prediction. Each value in the table reflects the average prediction error for a specific combination of input length and dataset size, with $k$ denoting the number of clusters used for the latitude and longitude components, respectively. Although extending the input segment $X$ provides the model with a longer historical context, this does not necessarily translate to improved predictive accuracy. The lowest average error was achieved with an input length of 32, which also corresponded to the largest dataset size (1107 samples). In contrast, increasing the input length to 40 and 48 time points led to a substantial reduction in the number of usable trajectories, and a corresponding increase in prediction error.

These findings suggest that the potential benefits of incorporating more extensive past information are counterbalanced by the negative effects of reduced training data. A smaller dataset may limit the model’s capacity to generalize across diverse trajectory patterns. Consequently, increasing the input length alone does not ensure better performance, particularly when it comes at the expense of dataset size.

\subsection{Qualitative Results}

\begin{figure}[hbt]
    \centerline{\includegraphics[width=1.0\columnwidth]{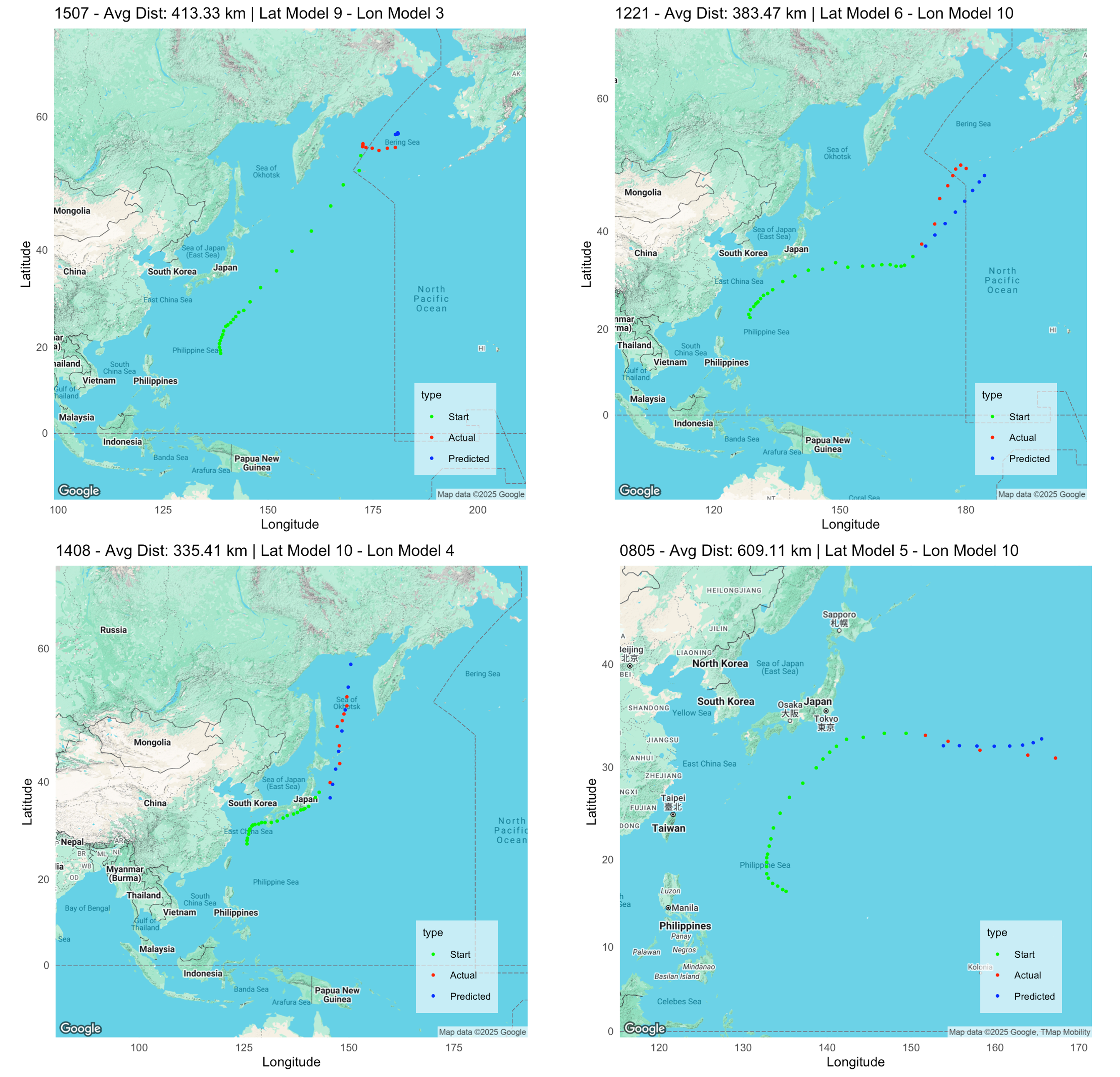}}\vspace{0.9ex}
    \caption{Results of regression without clustering against baseline model}
    \label{fig8}\vspace{1.5ex}
\end{figure}

To qualitatively assess the improvements achieved through clustering, several representative cases are shown in Figure 8. These trajectories had previously exhibited substantial prediction errors under the baseline model without clustering, primarily due to its tendency to regress toward the dominant northeastward path.

When re-evaluated using the clustering-based model, each case shows marked improvement. The predicted trajectories (blue) more closely follow the actual paths (red), capturing key directional changes that the baseline model failed to represent. This improvement is attributed to the use of cluster-specific regression models, which better reflect the structural characteristics of each trajectory group.

\section{Discussion and Conclusion}

This study introduced a clustering-based functional regression framework for predicting typhoon trajectories. The proposed method addresses the limitations of global regression models, which often fail to capture the diverse and irregular movement patterns of typhoons. By applying functional clustering separately to latitude and longitude trajectories, and training localized regression models within each cluster pair, the framework provides a more flexible and structure-aware approach.

Empirical results demonstrate that the clustering-based model significantly outperforms the baseline approach. The optimal configuration—consisting of 10 latitude clusters and 9 longitude clusters—reduced the average prediction error by more than 50\%. In addition to these quantitative improvements, qualitative examples showed that the model accurately captured atypical paths that the baseline method consistently mispredicted. These findings underscore the importance of accounting for trajectory-level heterogeneity in spatiotemporal modeling.

While the primary focus was on location-based prediction, the integration of auxiliary variables such as wind speed and central pressure did not lead to improved performance and, in some cases, degraded the results. This suggests that such variables may require more sophisticated treatment or modeling strategies. Furthermore, although reanalysis datasets could provide richer contextual information for typhoon systems, limited accessibility to these resources posed a constraint in the current study. Future work may explore more effective ways to incorporate meteorological covariates and leverage high-resolution reanalysis data to further enhance predictive accuracy.

\bibliography{refs}

\end{document}